\documentclass{article}
\usepackage{arxiv}
\usepackage[utf8]{inputenc} 
\usepackage[T1]{fontenc}    
\usepackage{dcolumn}
\usepackage{hyperref}       
\usepackage{url}            
\usepackage{booktabs}       
\usepackage{amsfonts}       
\usepackage{nicefrac}       
\usepackage{microtype}      
\usepackage{lipsum}
\usepackage{graphicx}
\graphicspath{ {./images/} }

\title{Self-acceleration and energy channeling in the saturation of the ion-sound instability in a bounded plasma}

\author{Liang Xu}

\author{
\hspace*{-1.2cm} Liang Xu \\
Department of Electrical Engineering and Information Science, Ruhr University Bochum,
D-44780 Bochum, Germany \\
\And
\hspace*{-0.9cm} Andrei Smolyakov and Salomon Janhunen \\
Department of Physics and Engineering Physics, University of Saskatchewan,
Saskatoon, Saskatchewan
S7N 5E2, Canada
\And
Igor Kaganovich \\
Plasma Physics Laboratory, Princeton University, Princeton, New Jersey 08543, USA
}

\begin{document}

\twocolumn[ 
  \begin{@twocolumnfalse} 
\maketitle
\begin{abstract}
  A novel regime of the saturation of the Pierce-type ion-sound instability in bounded ion-beam-plasma system is revealed in 1D PIC simulations. It is found that the saturation of the instability is mediated by the oscillating virtual anode potential structure. The periodically oscillating potential barrier separates the incoming beam ions into two groups. One component forms a supersonic beam which is accelerated to an energy exceeding the energy of the initial cold ion beam. The other component is organized  as a self-consistent phase space structure  of trapped ions with a wide  energy spread -- the ion hole. The effective temperature (energy spread) of the ions trapped in the hole is lower than the initial beam energy. In the final stage the ion hole expands over the whole system length. 
\end{abstract}
\end{@twocolumnfalse}
]

Plasma physics has yielded many important results in the theory of nonlinear waves. Earlier theories of coherent nonlinear waves, shocks and  solitons \cite{sagdeev69} have been expanded to include nonlinear structures in phase space, such as Bernstein-Green-Kruskal (BGK) states \cite{bernstein57}, electron and ion holes \cite{roberts1967nonlinear,Pecseli81,Schamel86,Eliasson06,kabantsev03,hasegawa82, Pecseli1984} and clumps\cite{Berk96,Berk97}. Such structures have been identified as the nonlinear saturated states of the beam-like kinetic instabilities in collisionless and  weakly collisional plasmas when the dynamics is dominated by a single coherent wave. The saturation mechanism via the formation of long lived phase space structures is a scenario alternative to the quasilinear saturation in the case of  a wide spectrum of overlapping modes resulting due to the flattening of the distribution function. Phase space holes/clumps have been detected in space plasmas \cite{Berk97,ergun01,temerin82,Bostrom88,cattell05} and shown to be important for the saturation of fast particle driven instabilities in fusion plasmas\cite{miyamoto80, Berk96,herrmann97,zonca14,chen16}. Many numerical experiments \cite{roberts1967nonlinear,Hutchinson17,Berk97,Pecseli1984,newman01,medvedev98,valentini11,lan19,hara19,zhou2016} have demonstrated phase space structures as a result of the saturation of bump-on-tail and Buneman instabilities. Most analytical and numerical studies have dealt with periodic systems. In practical laboratory applications, however, plasmas are typically bounded by walls that introduce important constraints. In this letter we present a novel mechanism of the nonlinear saturation of the ion-sound instability in a bounded plasma penetrated by a cold ion beam.  
 
In a warm infinite plasma the ion-sound instability of kinetic nature\cite{galeev1979review,wesson73} occurs when the current drift velocity exceeds the phase velocity of the ion sound waves $v_0>c_s/(1+k^2 \lambda_{De}^2)^{1/2}$, where $c_s$ is the ion sound velocity, $\lambda_{De}$ is the Debye length and $k$ is the wave number. In a bounded plasma of finite length there exists much stronger fluid instability of the Pierce type for $v_0<c_s$  which has been studied theoretically  \cite{Nakamura82,Kap15,Koshkarov15}, and has been observed in experiments\cite{Nakamura82}. The linear stage of the Pierce-type ion beam instability has been well studied theoretically.
Experiments in double plasma devices have revealed complex dynamics of Pierce instabilities induced by the ion and electron beams. Nakamura et. al. \cite{Nakamura82} have observed ion sound waves and nonlinear harmonics. The  oscillations were experimentally identified by Matsukuma et. al. \cite{Matsukuma01}. In the electron-beam-plasma system, Iizuka et. al. \cite{iizuka79, Iizuka85,Iizuka_nonlinear85} have demonstrated large scale current oscillations, the virtual cathode trapping ions and the double-layer formation. 

In this letter, we show that a strong ion sound instability induced by the ion flow in a bounded plasma saturates through the formation of the virtual anode potential structure that exhibits coherent large amplitude oscillations. The large amplitude periodic oscillations of this localized potential on one hand accelerate a fraction of the ion beam (initially subsonic) to supersonic velocity. On the other hand, the oscillating potential barrier stops a fraction of the ion beam, resulting in the formation of a long-lived large scale ion hole with a negative potential and a large population of trapped ions. Thus, a fraction of the initial energy of the ion beam $m_i n_{i0} v_0^2/2$ with $v_0<c_s$ is channeled to a higher energy creating a lower density supersonic beam with velocity $v_b>c_s$ , while rest of the beam energy is spread to lower energies forming the phase space vortex of trapped ions with a wide energy spread with an effective temperature $T_{tr} < m_i v_b^2$, where $m_i$ is the ion mass and $n_{i0}$ is the density of injected ions. In our case, $T_{tr} \simeq m_i v_b^2/6$.

A simplified 1D collisionless plasma model with a cold ion beam injected to the system is studied using the 1D3V electrostatic direct implicit particle-in-cell code (EDIPIC) \cite{sydorenko06}, which has been comprehensively validated and benchmarked \cite{Carlsson16,Xu17}. We focus only on the ion sound -type low frequency fluctuations $\omega/k \ll v_{th,e}$ where $v_{th,e}$ is the electron thermal velocity, so that Boltzmann-distributed electrons are used while the ions are computed fully kinetically. The ion and electron dynamics are coupled through the Poisson equation.

In the simulations, the grounded walls are set at locations $x=0$ and $x=d$, with the electric potential $\phi(x=0)=\phi(x=d)=0$. Ions with constant flux and velocity $v_0$ are injected from the left wall. The ions approaching the walls are absorbed. Initially, the electron density is set to be $n_{e0}=10^{14}m^{-3}$ and the (injected) ion density is slightly different to produce a charge perturbation with the value $\delta n=(n_{i0}-n_{e0})/n_{e0} \approx \pm 0.1 \%$. The sign of $\delta n$ was found to be important for the instability evolution. In our cases, positive initialization ($\delta n > 0$) was used leading to the ion hole formation. For simplicity, the ion mass $m_i$ is set to be 1 in atomic units. The Debye length in the simulations is calculated to be $\lambda_{De}=1.3 {\rm mm}$ and the simulation box length is set as $d=50\lambda_{De}$ with the spatial resolution of $\Delta x=\lambda_{De}/10$. Therefore, the simulation domain constitutes of $500$ cells and $d=6.5 {\rm cm}$.  Time step is chosen as $15\Delta t=\Delta x/v_{th,e}$ to fulfill the CFL condition during the simulation. Two thousand super-particles per cell are used. The electron temperature $T_e=3 {\rm eV}$ is used throughout this work.

\begin{figure}
\centering
\begin{tabular}{@{}cccc@{}}
\hspace*{-0.6cm}\includegraphics[width=0.31\textwidth]{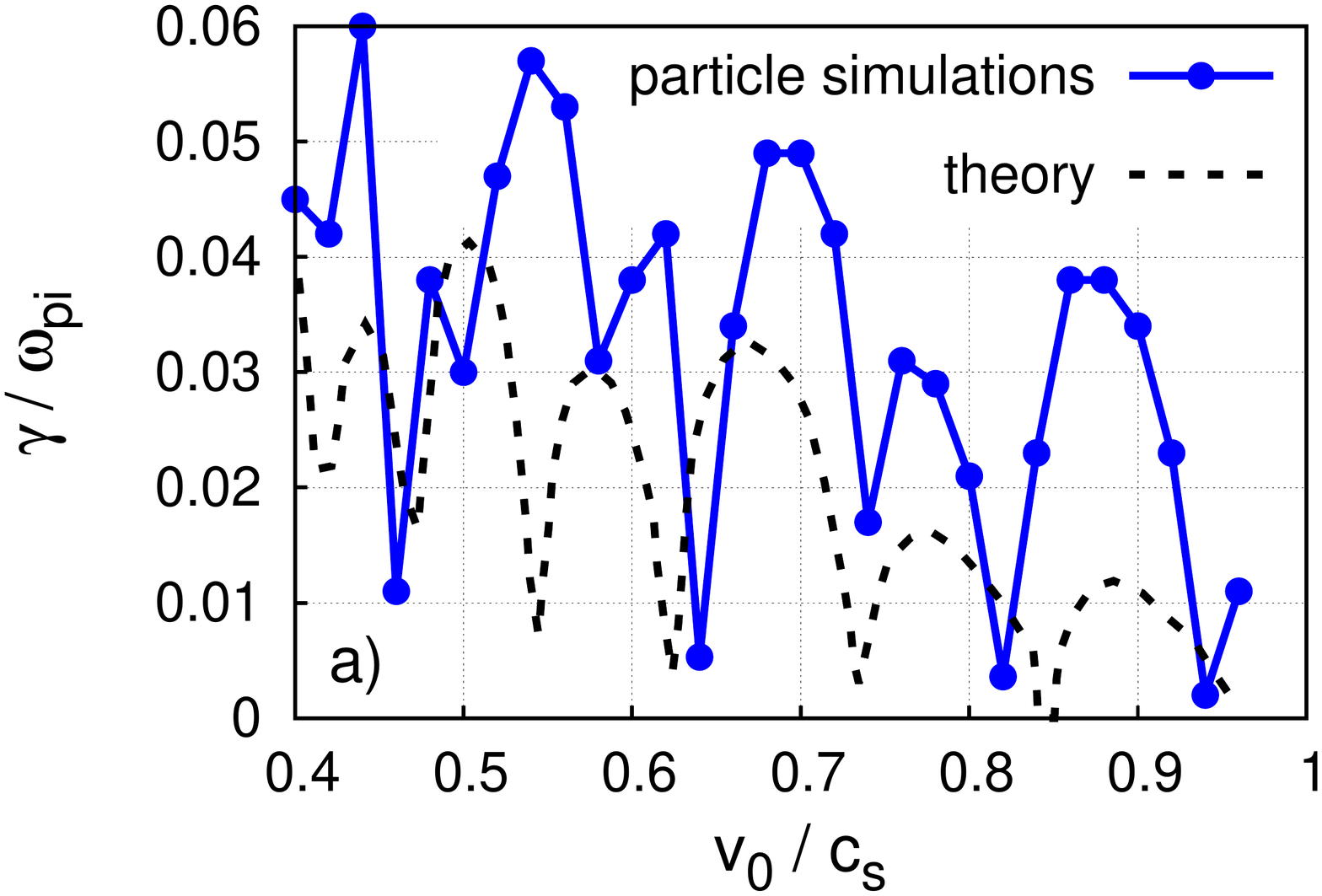}
\hspace*{-0.7cm}\includegraphics[width=0.31\textwidth]{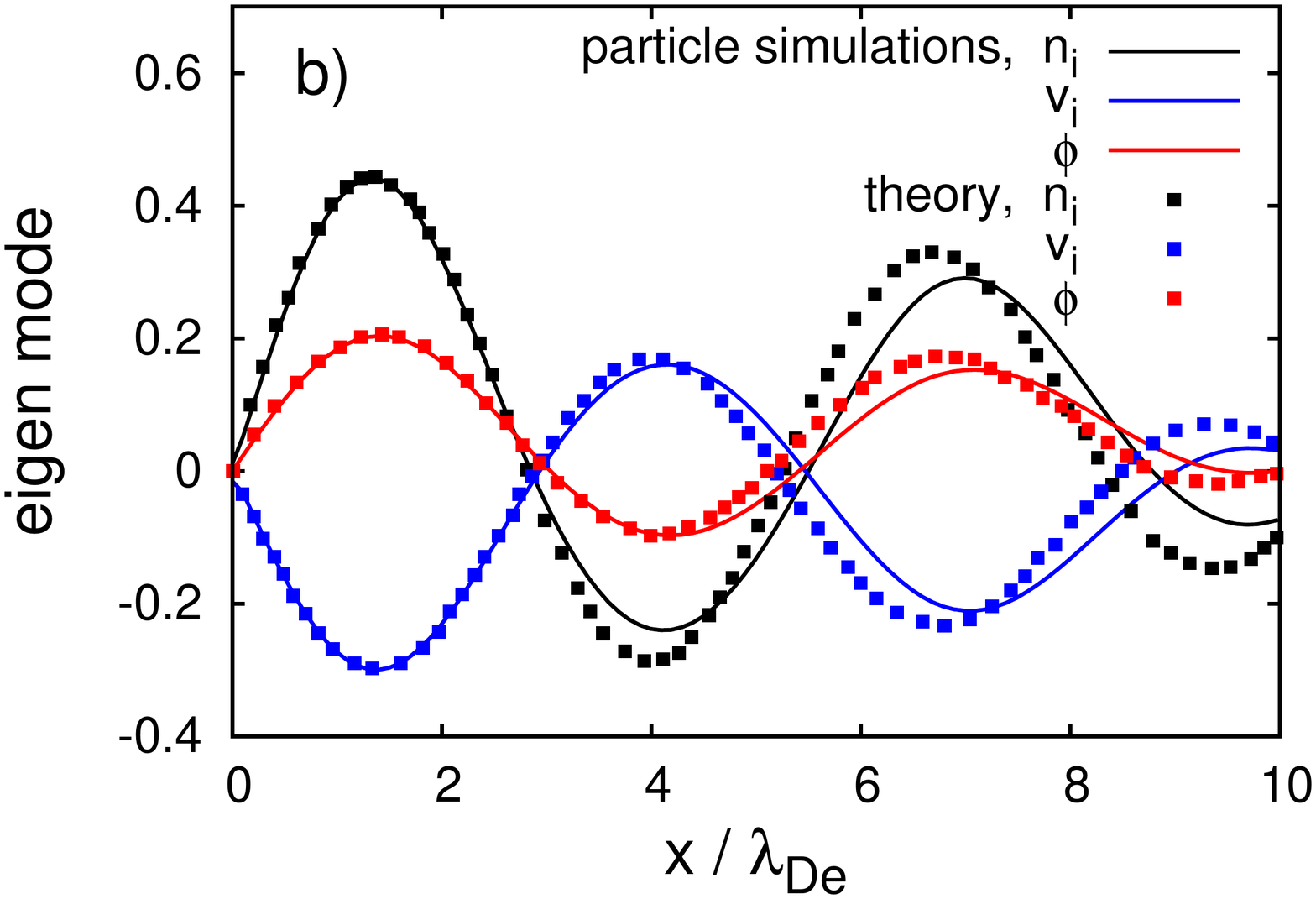}\\
\hspace*{-0.0cm}\includegraphics[width=0.28\textwidth]{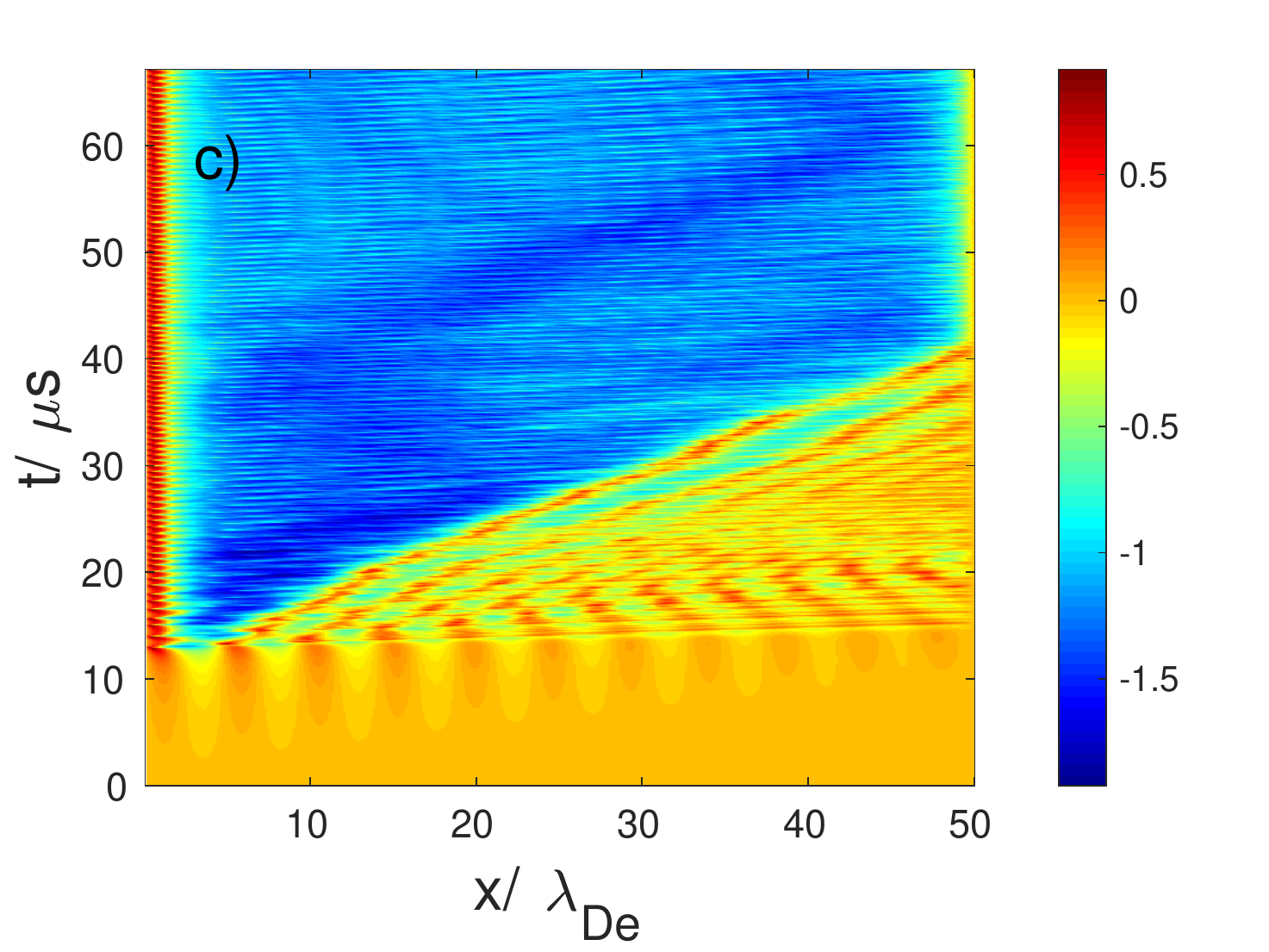}
\includegraphics[width=0.28\textwidth]{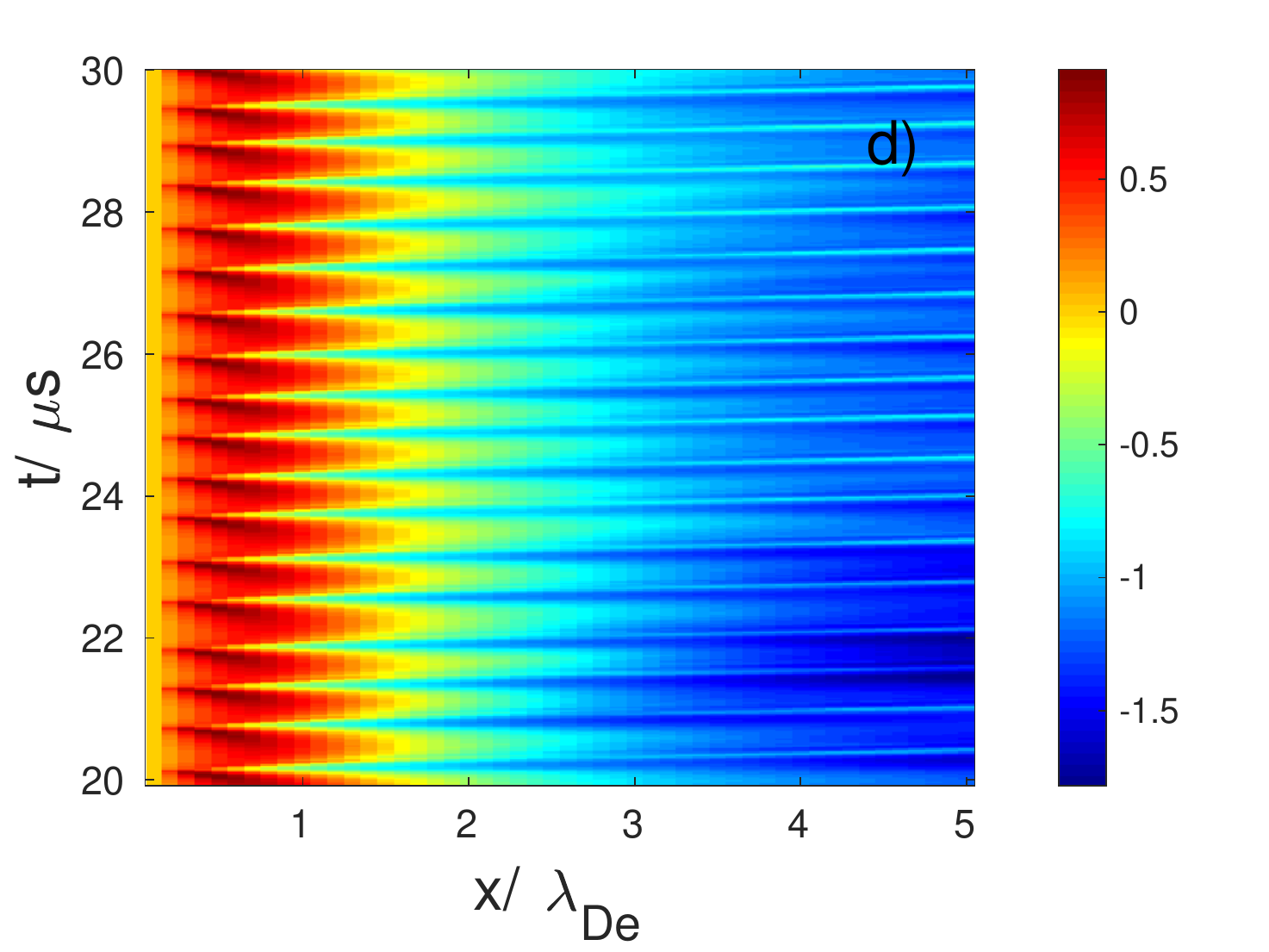}\\
\includegraphics[width=0.30\textwidth]{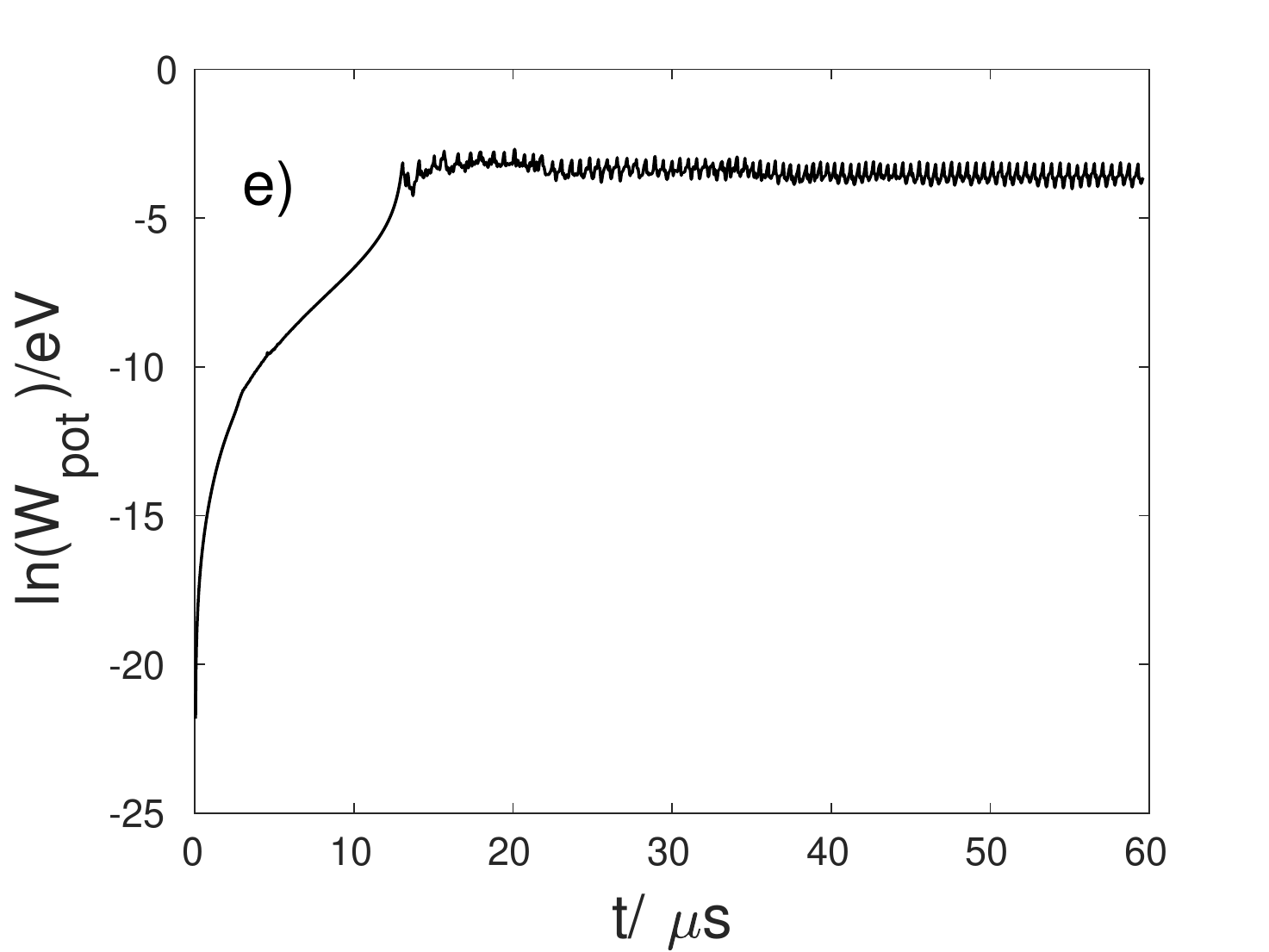}
\end{tabular}
\caption{ a) Linear growth rates and b) linear eigen modes of the ion sound instability from PIC simulations and fluid theory \cite{Koshkarov15}.  c) Potential oscillations in linear and nonlinear regimes.  d) Coherent oscillations of the virtual anode potential structure in the nonlinear regime. e) The evolution of the potential energy density as a function of time, showing the growth and saturation of the instability. }
\end{figure}

In the linear regime, when the kinetic effects of ion trapping are not important, our simulations recover the results of the analytical theory and the fluid simulations of the ion sound instability in a bounded plasma by  Koshkarov et. al. \cite{Koshkarov15}. 
This is shown in Figs. 1a and 1b which compare the growth rates and eigenmodes of the the Pierce type ion sound instability  obtained from our PIC simulations and the linear theory \cite{Koshkarov15}. In the simulations of  Figs. 1a and 1b, the system length is  $d=10 \lambda_{De}$, the ion beam velocity is in the range of $0.4-1.0c_s$ in Fig. 1a; and the ion beam velocity is $0.86 c_s$ in Fig. 1b. The linear growth rates and eigen-modes (ion density, velocity and potential) obtained from the simulations and theory,  Figs 1a and 1b, are in reasonable agreement taking into account that the analytical theory is formulated for the asymptotic limit of finite but small wave numbers, $k \lambda_{De} \ll 1$, while in our simulations $k \lambda_{De} \simeq 1$, see Fig. 1b. We note that according to the theory, the ion sound instability occurs only for $v_0<c_s$, while the case $v_0>c_s$ is stable for this type of instability.

\begin{figure}[htp]
\vspace*{-3.5cm}\hspace*{-5.8cm}\includegraphics[width=1.1\textwidth]{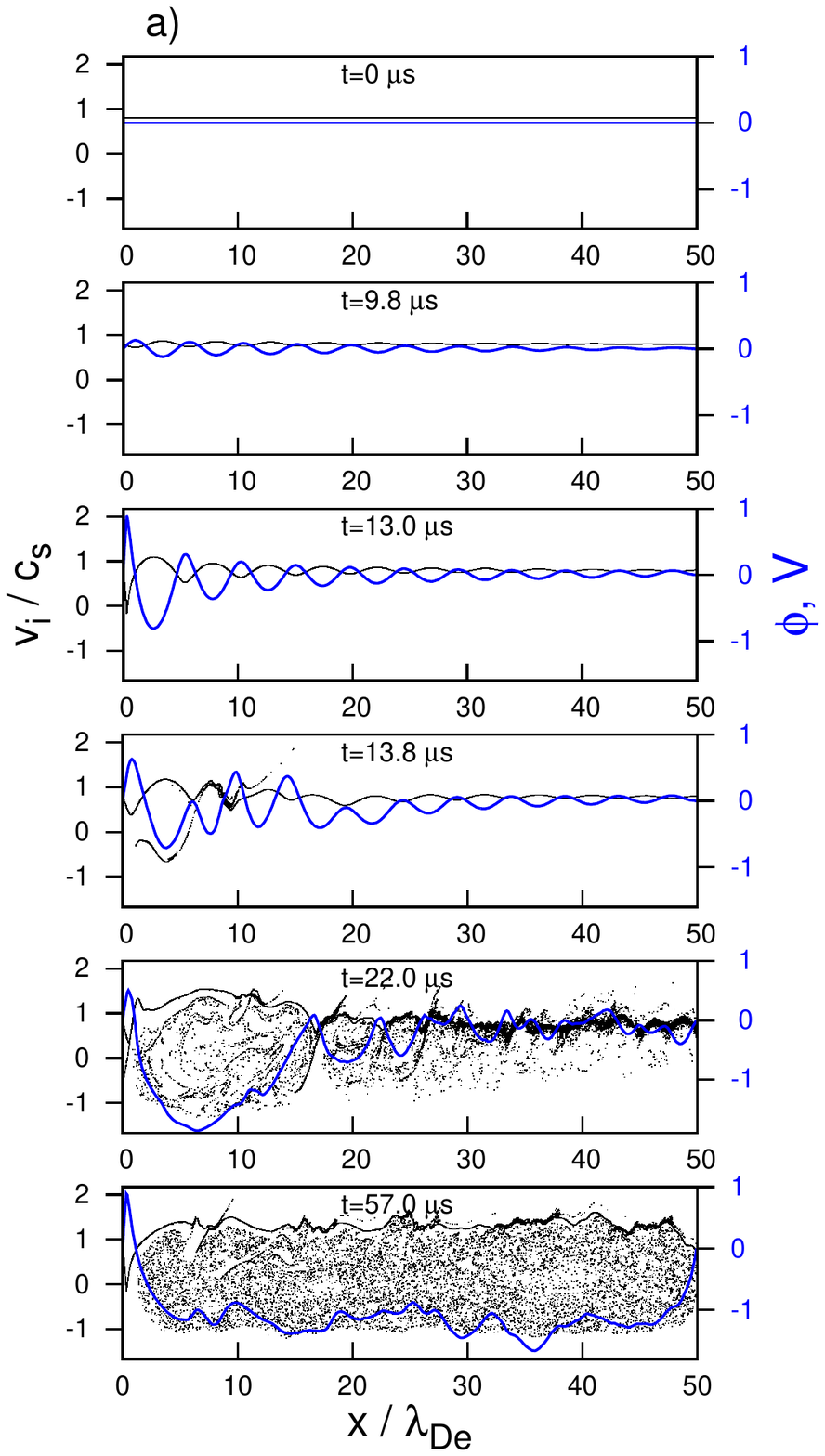}
\hspace*{-1.0cm}\includegraphics[width=0.45\textwidth]{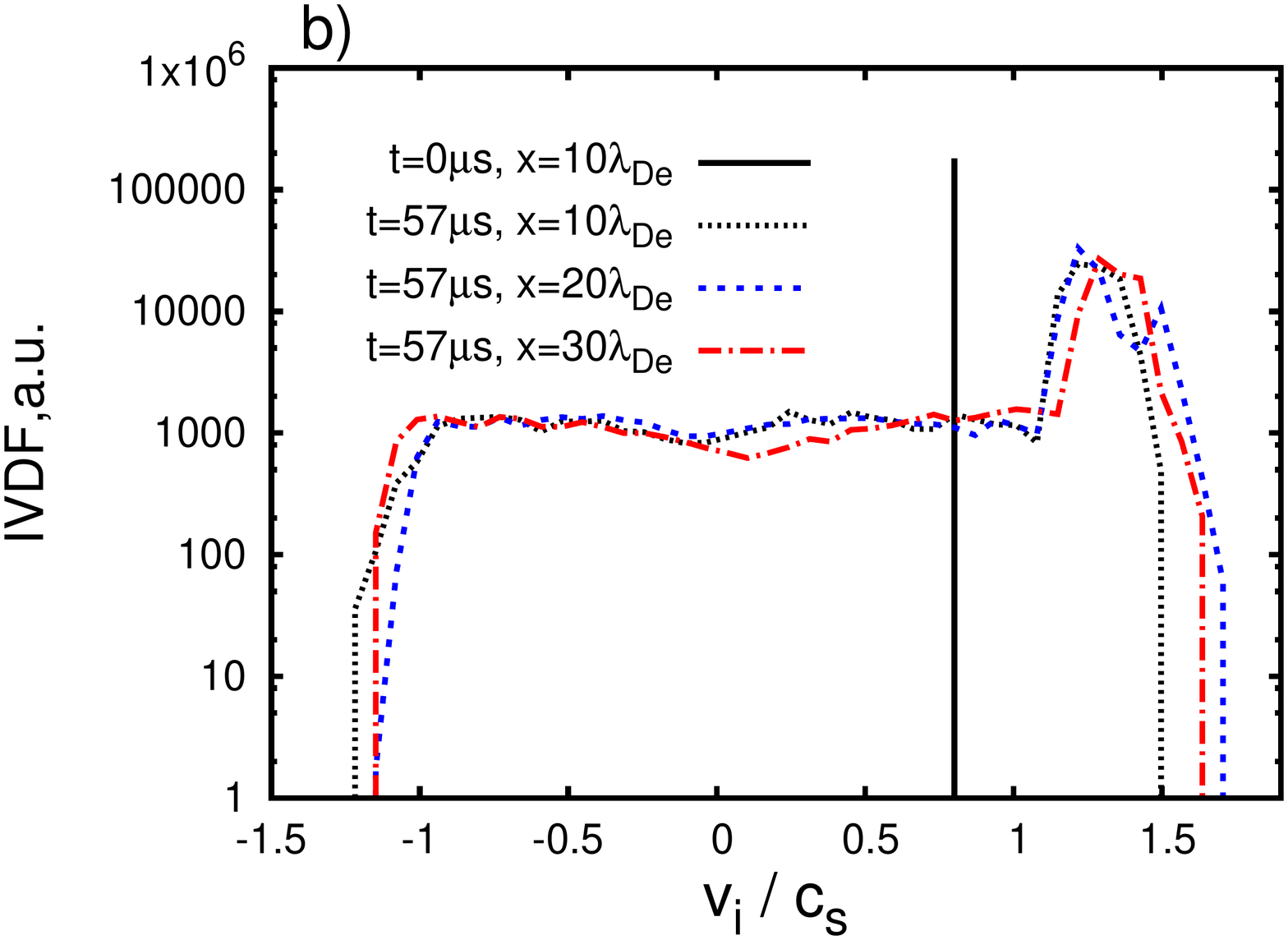}

\caption{a) The ion phase plane and potential profiles at different times of the linear instability ($t<13 \mu s$), wave breaking forming multi-valued solutions ($t>13.8 \mu s$) to the large-scale ion hole of the system size in the final stage $t=57 \mu s$ b) The ion velocity distribution in the initial and final stages.}
\end{figure}

As was shown analytically \cite{Koshkarov15}, the instability occurs in the alternating  oscillatory (real part of frequency $Re(\omega) \neq 0$) and aperiodic ($Re(\omega) = 0$) zones as the ion beam velocity changes. In our nonlinear simulations, Figs. 1c, 1d and 1e, the ion beam velocity is $v_0=0.8 c_s$. With  this value, and  $d=50\lambda_{De}$, the  instability occurs in the aperiodic zone, so the fluctuations grow as a standing wave as it is clearly seen in Fig. 1c showing the spatial and temporal evolution of the potential. Fig. 1e shows the time evolution of the normalized potential energy density $W_{pot}=1/(n_{i0}d)\int_0^d{1/2 \epsilon_0 E^2}dx$, from which the growth rate is calculated to be about $0.02 \omega_{pi}$ in the linear stage, consistent with the fluid numerical calculation of the linear theory. 

The linear growth and the transition to the nonlinear stage can be seen in the ion phase space plane and potential profile shown in Fig. 2a for different points in time. Initially, the potential $\phi (x) = 0$ with a mono-energetic ion flow throughout the simulation box. The convective mode amplification starts at the left wall, as predicted by the linear theory. The exponential growth occurs after the ion transit (over the box length) time, and the potential reaches $0.1 V$ near the left wall around $9.8 \mu s$. The ion velocity is still single valued at this time, which suggests that the fluid description still holds. 
Around $t\simeq 13.0 \mu s$, the potential perturbation forms a peaked structure (in the first half-wave-length)  ending in the singularity  and subsequent multi-valued solutions, seen at time $t=13.8 \mu s$ in Fig. 2a.   Multi-valued solutions occur due to the ion reflection by the potential barrier structure, as illustrated  in Fig. 2a by the appearance of the  ions with  negative velocities. The potential barrier near the left wall (called below as a virtual anode) oscillates in both time and space; the zoomed-in space-time profile of the potential near the left wall is shown in Fig. 1d. The virtual anode oscillations are rather coherent in time with a frequency of the order $\omega_{pi}$ corresponding to the ion sound oscillations in the short wavelength regime, $k \lambda_{De} \gg 1$.

The periodically oscillating potential barrier (oscillating virtual anode) creates two related phenomena. In the raising stage, it accelerates the beam ions to a velocity larger than the velocity of the initial beam, i.e., $v_b>v_0$ (in this case $v_0=0.8 c_s$).  In fact, the fraction of the original beam ions is accelerated to supersonic velocities of the order of $1.3 c_s$, see Fig 2b. On the other hand, the large potential barrier breaks the flow of ions, decelerating and partially reflecting  them. In the decaying stage of the potential, some decelerated ions are released forming the population of the trapped ions of  a finite temperature that result in the ion hole. The formation of the ion hole is clearly seen at $t=22 \mu s$, Fig 2a. The trapped ions have a finite temperature due to scattering from the periodically oscillating virtual anode. Therefore oscillating virtual anode structure results in the acceleration of some ions forming a supersonic beam with $v_b>c_s>v_0$ and the heating of other ions. In Fig. 2b, this energy channeling is shown by the evolution  of the ion distribution from the delta function of the cold ions beam with $v_i=v_0=0.8 c_s$ at $t=0\mu s$ to the wide distribution with a beam fraction of the order $v_i=v_b=1.3 c_s$ and the trapped ion population with a finite temperature (effective mean energy) $T_{tr} \simeq 0.8 eV$ at $t=57 \mu s$. The trapped ions result in the ion hole with negative potential, which eventually expands through the whole system as shown in Fig 2a. 

\begin{figure}[htp]
\centering
\vspace{-1cm}\hspace*{-1cm}\includegraphics[width=0.6\textwidth]{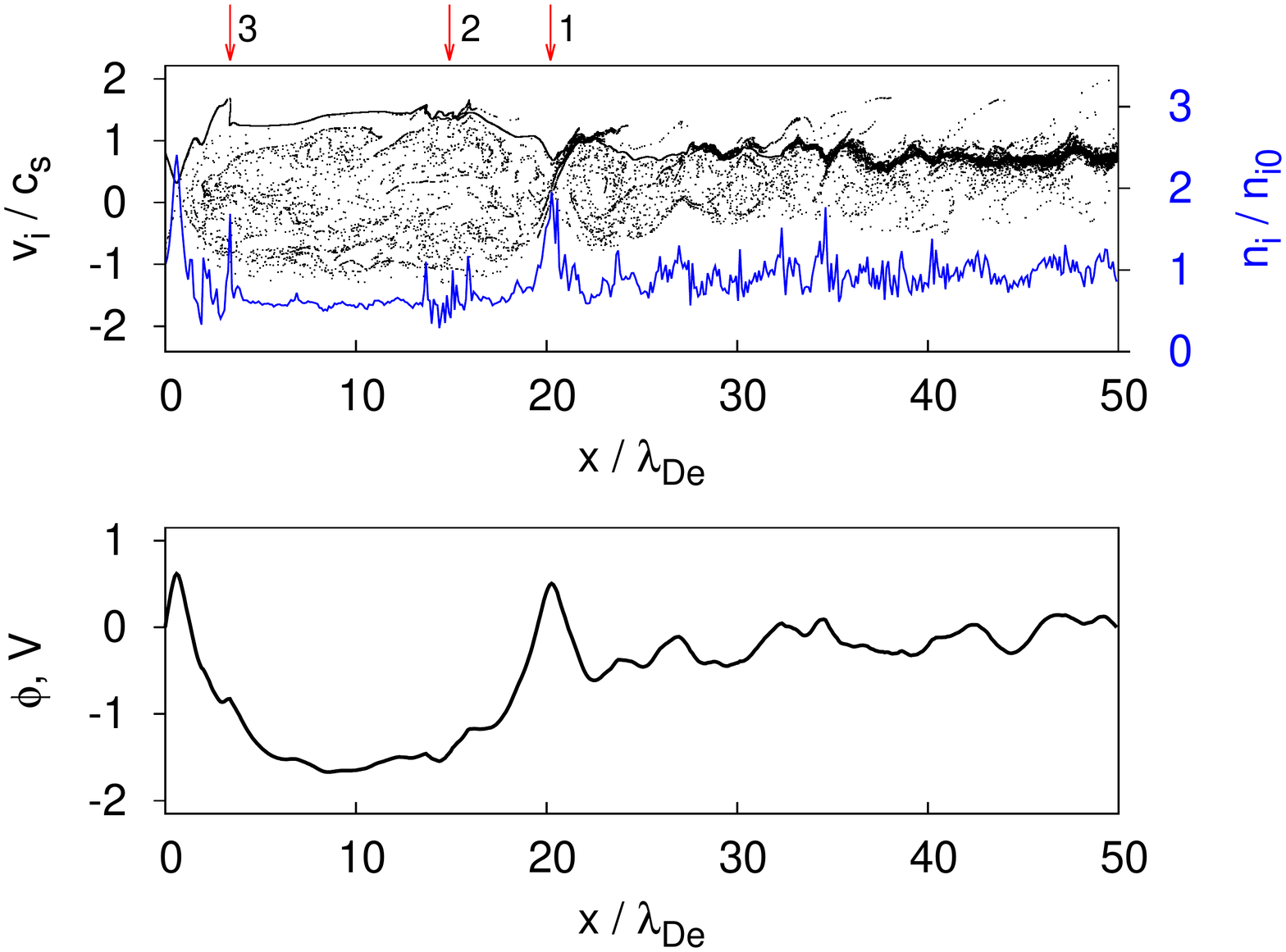}
\caption{Several decelerating/scattering events with enhanced density bunches (labeled as 1,2,and 3)  are shown in the ion phase space plane, ion density, and potential profile at the snapshot $t=23.93 \mu s$.}
\end{figure} 

The ion trapping occurs in several stages as a result of consecutive decelerating and scattering events. Several of such events are shown in the ion phase plane, ion density and potential profiles in one snapshot $t=23.93 \mu s$ as  shown in Fig. 3. One can see the localized  bunches (labeled as 1,2, and 3) of increased ion density at the locations of decreased ion velocity, which also correlate with the peaks in the  potential. The most recent bunch of decelerated ions (the arrow 3 position) is shown around the virtual anode location  near the left boundary. The bunch which was released earlier is scattered at the location marked by the arrow 2. The earliest bunch in this snapshot is marked by the arrow 1, where one can see the velocity depletion and strong density enhancement  at the location of large potential peak. Effectively this location corresponds to the ion hole boundary that moves toward the right wall. The velocity of the ion hole expansion is estimated here as the thermal (mean) velocity of trapped ions $<v_{tr}>=\frac{1}{2}v_b$, and the temperature of the trapped ions is $T_{tr}=\frac{1}{6}m_i v_b^2$.  

\begin{figure}[htp]
\hspace*{1cm}\includegraphics[width=0.4\textwidth]{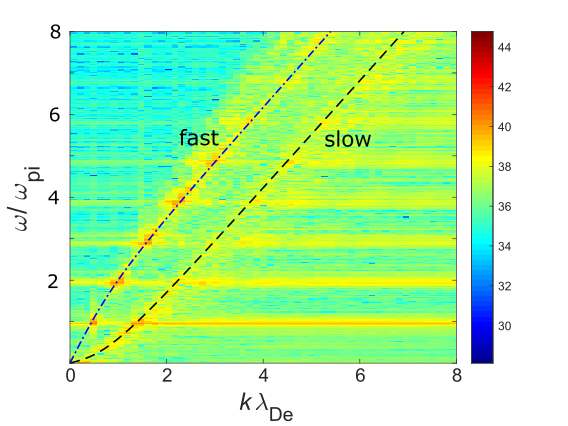}
\caption{A Fourier transform of the ion density over the whole simulation box and in the time range $45-70\mu s$ shows the the spectra of fluctuations. The two lines are the theoretical prediction given by Eq. 1.} 
\end{figure}

In the saturated stage, the fluctuation spectrum consists of the Doppler shifted ion sound modes and the nonlinear harmonics of the short wavelength ion sound $\omega=\omega_{pi}$ for $k \lambda_{De}\gg 1$, as illustrated in Fig. 4 for the Fourier transform of ion density over the whole space and in the time range $45-70 \mu s$. The two lines in Fig. 4 are the dispersion relations of Doppler-shifted ion sound waves:

\begin{equation}
    \omega/\omega_{pi} =  k v_{b}/\omega_{pi} \pm k \lambda_{De}[1+(k \lambda_{De})^2]^{-1/2}
\end{equation}

The sign $\pm$ represents the two branches of the ion sound wave, fast beam mode and slow beam mode. As mentioned in Fig. 2b, the beam velocity $v_{b} \approx 1.3 c_s$. It is shown in Fig. 4 that calculation by Eq. 1 agrees well with the simulation. In addition, one can identify the oscillations with ion plasma frequency and its multiple harmonics, which are the typical feature of oscillating virtual anode \cite{piel96}, also reminiscent of the  experimental observations in Ref. [29].
  
In summary: we report a novel mechanism of trapped ion phase space hole formation in a bounded ion-beam-plasma system driven by the Pierce-type ion sound instability due to the presence of boundaries. Our simulations agree well with the linear theory of such an instability. In the nonlinear stage we observe an oscillating potential barrier structure (i.e., the virtual anode) splitting the ion beam and channeling energy to a population of the  higher energy ions while  heating the ions at lower energies. The higher energy fraction of accelerated ions forms the supersonic ion beam, while the group of decelerated ions forms a system-long ion hole of trapped ions.  Similar nonlinear phenomena associated with the oscillating virtual anode may be expected in Hall plasmas \cite{Kap15}, Q machines \cite{kuhn1984} and material processing devices \cite{xu08,chen19}, as well as the bounded ion beam-plasma systems as studied in double plasma devices \cite{Nakamura82} and inertial confinement fusion drivers \cite{poukey1981,lemons1982,schamel93} when $\omega_{pi}d/v_0>\pi$.

This work was supported in part by NSERC Canada and the Air Force Office of Scientific Research FA9550-15-1-0226. Computational resources were provided by  ComputeCanada/WestGrid. The authors would like to express their gratitude to Ralf Peter Brinkmann,
Edward Startsev and Jian Chen for insightful discussions and suggestions.




\end{document}